\documentclass[10pt, conference, letterpaper]{IEEEtran}
\usepackage{amsmath,amsthm}
\usepackage{tabularx,multirow}
\usepackage{graphicx,subfigure,comment}
\usepackage[table]{xcolor}
\usepackage{url} 
\usepackage{dsfont}
\usepackage{cite}
\usepackage{siunitx}
\usepackage{physics} 
\usepackage{pstool}
\usepackage{enumitem}
\usepackage{autonum} 

\setlist[itemize]{leftmargin=*}
\setlist[enumerate]{leftmargin=*}

\newcommand{\Fig}[1]{Fig.~\ref{fig:#1}}
\newcommand{\Prop}[1]{Property~\ref{prop:#1}}
\newcommand{\Ex}[1]{Example~\ref{ex:#1}}

\newcommand{\Sec}[1]{Sec.~\ref{sec:#1}}
\newcommand{\Tab}[1]{Tab.~\ref{tab:#1}}
\newcommand{\Eq}[1]{(\ref{eq:#1})}

\newcommand{\ind}[1]{\mathds{1}_{#1}}

\newcommand{\Vc}{\mathcal{V}}

\newcommand{\Mc}{\mathcal{M}}
\newcommand{\Sc}{\mathcal{S}}
\newcommand{\PP}{\mathds{P}} 

\newtheorem{theorem}{Theorem}
\newtheorem{property}{Property}

\newtheorem{example}{Example}
\newtheorem*{example*}{Example}

\allowdisplaybreaks

\begin{document}

\title{
Getting the Most Out of Your VNFs:\\
Flexible Assignment of Service Priorities in 5G
} 

\author{\IEEEauthorblockN{Francesco Malandrino$^{\ast,\dagger}$, Carla-Fabiana Chiasserini$^{\dagger,\ast}$}
\IEEEauthorblockA{$\ast$: CNR-IEIIT, Torino, Italy $\quad\quad$ $\dagger$: Politecnico di Torino, Torino, Italy \vspace*{-1cm}}
} 

\maketitle

\begin{abstract}
Through their computational and forwarding capabilities, 5G networks can support multiple vertical services. Such services  may include several common virtual (network) functions (VNFs), which could be shared to increase resource efficiency. In this paper, we focus on the seldom studied VNF-sharing problem, and decide (i) whether sharing a VNF instance is possible/beneficial or not, (ii)
how to {\em scale} virtual machines hosting the VNFs to share,
and (iii) the priorities of the different services sharing the same VNF. These decisions are made with the aim to minimize the mobile operator's costs while meeting the verticals' performance requirements. Importantly, we show that the aforementioned priorities should not be determined {\em a priori} on a per-service basis, rather  they should change across VNFs since such additional flexibility allows for more efficient solutions. We then present an effective methodology called FlexShare, enabling near-optimal VNF-sharing decisions in polynomial time. Our performance evaluation, using real-world VNF graphs, confirms the effectiveness of our approach, which consistently outperforms baseline solutions using per-service priorities.
\end{abstract}

\section{Introduction}
\label{sec:intro}

5G networks differ from their previous counterparts in several key features. Among them, two especially relevant aspects are the computational capabilities with which networks are endowed, and the relationship between network operators and vertical industries (e.g., multimedia content providers, automotive industries, smart factories).

Thanks to software-defined networking (SDN) and network function virtualization (NFV), cellular networks have now the  ability to {\em process} data, as well as to forward them. A set of hosts, be them physical servers or virtual machines, run virtual network functions (VNFs), performing network-related (e.g., firewalls), or service-specific (e.g., video transcoding) tasks. Such an arrangement is beneficial for both mobile network operators (MNOs), who can fully utilize their infrastructure, and verticals, who can use it to deploy their services and enjoy lower delays.

The relationship between MNOs and verticals is evolving accordingly. Verticals can enter business relations with MNOs, using their network infrastructure to provide  services with the required quality-of-service level. Indeed, vertical services are specified as a set of interconnected VNFs, along with {\em per-service} target key performance indicators (KPIs), e.g., throughput, delay, or reliability.  
Supporting services of  multiple verticals, with different target KPIs, through the same MNO infrastructure, is possible thanks to {\em network slicing}.  This is a powerful concept  enabling multiple logical networks as independent business operations, on a common physical infrastructure~\cite{slicing}. Notably, network slicing also accounts for composed services, i.e., service VNF graphs including sub-graphs, each representing a child service~\cite{slicing2}. Indeed, different network slices may contain one or more common sub-slices~\cite{5gppp-architecture,ietf-mano}, a typical example being the evolved packet core (EPC) of cellular  networks~\cite{slicing2}.

When creating a network slice, the MNO is in charge of assigning to it the needed resources (e.g., virtual machines and links connecting them), and of deciding which VNFs each host should run.  This problem, known as VNF placement, pursues  the twofold objectives of (i) ensuring that the target KPIs are met, and (ii) minimizing the cost for the MNO. Importantly, the latter can be achieved by {\em sharing} individual VNFs or sub-slices, among multiple services whenever possible. 

The vast majority of studies on VNF placement~\cite{placement-cinesi,placement-infocom,noi-infocom18} implicitly assume that (i) all placement decisions are made by one entity, typically the NFV Orchestrator (NFVO) of the Management and Network Orchestration (MANO) framework~\cite{etsimano,ietf-mano}, and (ii) such entity makes fine-grained decisions about how individual hosts and links are used. However, such a behavior is not the only one included in standards, and is not typical of real-world 5G implementations. Indeed, ETSI IFA~007 specifies four possible granularity levels for placement decisions, namely, Point of Presence  (PoP) -- e.g., a data center --, zone\footnote{{\em Zone} refers to a subset of hosts within a PoP.} group, zone, and individual host, with real-world 5G implementation efforts envisioning the NFVO to make PoP-level decisions \cite{pimrc-wp4,norma}. 

\begin{figure}
\includegraphics[width=1\columnwidth]{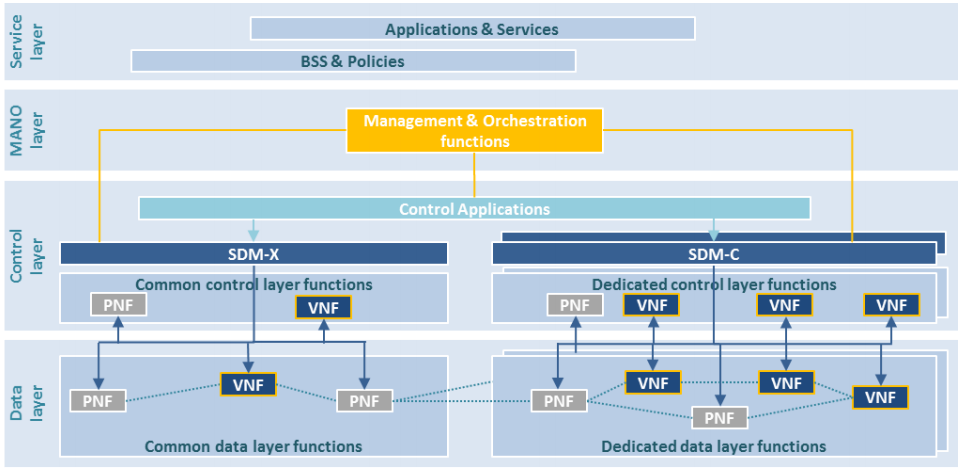}
\caption{
   Architectural view of 5G networks according to 5G-PPP. Source:~\cite{slicing2}.
    \label{fig:5gppp}
} 
\vspace{-5mm}
\end{figure}

Fine-grained decisions on sharing VNF instances within individual PoPs are made by other, non-MANO entities, as envisioned by IETF~\cite[Sec.~3]{ietf-mano}, the NGMN alliance~\cite[Sec.~8.9]{ngmn-mano}, and 5G-PPP~\cite[Sec.~2.2.2]{5gppp-architecture}. For sake of concreteness, we take as a reference the latter architecture, which includes an entity called Software-Defined Mobile Network Coordinator (SDM-X). As summarized in \Fig{5gppp}, the SDM-X works at a lower abstraction level than the MANO entities, and is in charge of solving the {\em VNF-sharing} problem, managing the VNFs that are common to multiple services.

\begin{figure}
\psfragfig[width=1\columnwidth]{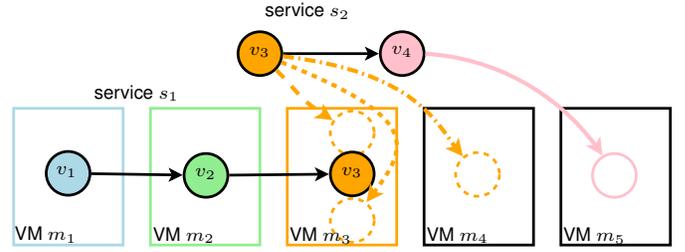}{
\psfrag{s1}[l][m]{\scriptsize{{\sf service~$s_1$}}}
\psfrag{s2}[l][m]{\scriptsize{{\sf service~$s_2$}}}
\psfrag{h1}[l][b]{\scriptsize{{\sf VM~$m_1$}}}
\psfrag{h2}[l][b]{\scriptsize{{\sf VM~$m_2$}}}
\psfrag{h3}[l][b]{\scriptsize{{\sf VM~$m_3$}}}
\psfrag{h4}[l][b]{\scriptsize{{\sf VM~$m_4$}}}
\psfrag{h5}[l][b]{\scriptsize{{\sf VM~$m_5$}}}
\psfrag{v1}[c][m]{\scriptsize{{\sf $v_1$}}}
\psfrag{v2}[c][m]{\scriptsize{{\sf $v_2$}}}
\psfrag{v3}[c][m]{\scriptsize{{\sf $v_3$}}}
\psfrag{v4}[c][m]{\scriptsize{{\sf $v_4$}}}
} 
\caption{
    Example of the VNF-sharing problem. A PoP is serving service~$s_1$, with VNFs~$v_1$--$v_3$ deployed at VMs~$m_1$--$m_3$. It is then requested to deploy~$s_2$, using VNFs~$v_3$ and~$v_4$. No isolation is requested, so services can share VNFs if convenient. For~$v_4$, the only option is devoting an unused VM to it, $m_5$~in the example (pink line). For~$v_3$, instead, there are three options: re-using the instance of~$v_3$ at~$m_3$, giving~$s_2$ a lower priority than~$s_1$ (dashed line); doing the same but giving~$s_2$ a higher priority (dotted line); devoting~$m_4$ -- currently unused -- to~$v_3$ (dash-dotted line), thus having two VMs running~$v_3$.
    \label{fig:idea}
} 
\vspace{-5mm}
\end{figure}

This is the problem that, differently from traditional VNF placement studies, we address in this paper. Specifically, we solve the VNF-sharing problem within a PoP, by deciding,  for each newly requested service,
\begin{itemize}
    \item whether it is possible and convenient to re-use  the existing VNF instances\footnote{For simplicity, and without loss of generality, we will refer to common VNFs only, instead of common VNFs and sub-slices.};
    \item which priority to give to services sharing the same  VNF, in order to meet the target KPIs;
    \item whether and to which extent to {\em scale up} the computational capability of virtual machines (VMs) within the PoP.
\end{itemize}
A simple instance of the VNF-sharing problem we address is presented in \Fig{idea}. For each VNF of the newly-requested service~$s_2$, we have to decide which VM shall run it (this decision is trivial for~$v_4$, for which no instance exists yet) and the priority to assign to requests, e.g., REST queries, of each service in each shared VNF.

{\bf Contributions}. 
To effectively tackle the VNF-sharing problem, we make the following main contributions:
\begin{itemize}
    \item we  observe that allowing {\em flexible} priorities for each VNF and service,  makes it possible to meet KPI targets at a lower cost for the MNO;
    \item based on  the above key observation, we present a system model capturing all the relevant aspects of the VNF-sharing problem and the entities it involves, including the capacity-scaling and priority-setting decisions it requires;
    \item leveraging convex optimization, we devise an efficient, integrated solution methodology called FlexShare, able to make swift, high-quality decisions concerning VM usage, priority assignment, and capability scaling;
    \item we evaluate the computational complexity of FlexShare and study its performance against real-world VNF graphs.
\end{itemize}

In the rest of the paper, we begin by presenting an example motivating the need for flexible priorities across VNFs (\Sec{example}). Then, \Sec{model} introduces our system model and problem formulation, while \Sec{algo} describes the FlexShare solution strategy. Our reference scenarios and numerical results are discussed in \Sec{results}. Finally, after reviewing related work in \Sec{relwork}, \Sec{conclusion} concludes the paper.

\section{The role of priorities}
\label{sec:example}

Before addressing the problem of whether it is convenient to share a VNF among multiple services or not, let us highlight the role of priorities while sharing a VNF instance. Three main approaches can be adopted for VNF sharing:
\begin{itemize}
    \item {\em per-service} priority, associated with each service and constant across different VNFs;
    \item {\em per-VNF} priority, associated with each service and VNF, thus, given a service, 
it may vary across different VNFs;
    \item {\em per-request} priority, associated with individual service requests, e.g., REST queries, it may vary across the different VNFs on a per-request basis.
\end{itemize}
In the following example, we focus on the first two steps of the above flexibility ladder and show  how higher flexibility in priority assignment is associated with a higher efficiency in handling  service traffic. 

\noindent\rule{1\columnwidth}{0.5mm}
\begin{example}[The importance of flexible priorities]
\label{ex:prios}
Consider the two services, $s_1$ and $s_2$, depicted in \Fig{example}, requested by a vertical specialized in video surveillance systems. $s_2$  includes two VNFs executing transcoding and motion detection, respectively, while   $s_1$ is composed of $s_2$ and a VNF performing face recognition. Each VNF should run in its own VM, and network transfer times between VMs are neglected. Adopting a well-established and convenient approach~\cite{placement-infocom,noi-infocom18}, let us model 
VNFs as M/M/1 traffic queues processing service requests, and the services as queuing chains, with arrival rates~$\lambda_1=\SI{2}{requests/\milli\second}$ and~$\lambda_2=\SI{1}{requests/\milli\second}$, respectively.
Also, consider service delay as the main performance metric and let the target average delay be~$D^{\max}_1=D^{\max}_2=\SI{1.1}{\milli\second}$ for both services. Then assume that, given the allocated computation resources,  the service rate of the transcoding and motion detection  is~$\mu_\text{tc}=\mu_\text{md}=\mu=\SI{5}{requests/\milli\second}$, while that of face recognition is $\mu_\text{fr}=\SI{9.15}{requests/\milli\second}$.

To meet the delay targets, $s_2$~requests must traverse the transcoding and motion detection with a combined sojourn time of \SI{1.1}{\milli\second}, while $s_1$~requests must do the same in at most \SI{0.96}{\milli\second} (i.e., the target average delay~$D^{\max}_1$ minus the sojourn time at the face recognition VNF of $\frac{1}{\mu_\text{fr}-\lambda_1}=\SI{0.14}{\milli\second}$). We now show that there is no way of setting {\em per-service} priorities that allow this.

\noindent{\em Case 1: Higher priority to~$s_1$.} This choice would make intuitive sense since $s_1$~requests have to go through more processing stages than $s_2$ requests, within the same deadline. 
In this case, $s_1$~requests incur a sojourn time of~$\frac{1}{\mu-\lambda_1}=\frac{1}{5-2}=\SI{0.33}{\milli\second}$ for each of the common VNFs, resulting in a total delay $D_1=\SI{0.8}{\milli\second}$, well within the target. However, the sojourn time of $s_2$~requests at each of the shared VNFs becomes~\cite[Sec.~3.2]{kleinrock-vol2}:
$\frac{1/\mu}{\left(1-\frac{\lambda_1}{\mu}\right)\left(1-\frac{\lambda_1+\lambda_2}{\mu}\right)}
=
\frac{1/5}{\left(1-\frac{2}{5}\right)\left(1-\frac{1+2}{5}\right)}\approx\SI{0.83}{\milli\second}$,
which results in a total delay of~$D_2\approx\SI{1.66}{\milli\second}>D^{\max}_2$.

\noindent{\em Case 2: Higher priority to~$s_2$.} It is easy to verify that giving higher priority to~$s_2$ implies that $s_1$~misses its target delay.

\noindent{\em Case 3: Equal priority.} Giving the same priority to both services results in a sojourn time of $\frac{1}{\mu-\lambda_1-\lambda_2}=\frac{1}{5-1-2}=\SI{0.5}{\milli\second}$ for each of the common VNFs, and in a total delay of~$D_2=\SI{1}{\milli\second}<D^{\max}_2$ and~$D_1=\SI{1}{\milli\second}+\SI{0.14}{\milli\second}>D^{\max}_1$.

\begin{figure}
\centering
\psfragfig[width=1\columnwidth]{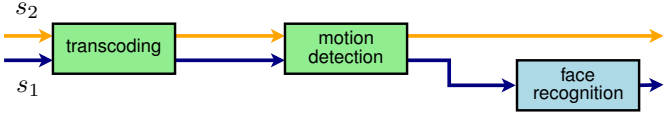}{
\psfrag{s2}[c][b]{$s_1$}
\psfrag{s1}[c][m]{$s_2$}
\psfrag{trans}[c][m]{\scriptsize{{\sf transcoding}}}
\psfrag{mot}[c][m]{\scriptsize{{\sf motion}}}
\psfrag{det}[c][m]{\scriptsize{{\sf detection}}}
\psfrag{face}[c][m]{\scriptsize{{\sf face}}}
\psfrag{rec}[c][m]{\scriptsize{{\sf recognition}}}
} 
\caption{\Ex{prios}: two video surveillance services,~$s_1$ and~$s_2$, with $s_2$ including performing transcoding and motion detection VNFs, and $s_1$ composed of $s_2$ and an additional face recognition stage.
    \label{fig:example}
} 
\vspace{-5mm}

\end{figure}

\noindent{\em Flexible priorities.} 
Assume that $s_1$ and $s_2$ have priority in the transcoding and in the motion detection VNF, respectively. Then, $D_1 \mathord{=}\frac{1}{\mu-\lambda_1}\mathord{+}\frac{1/\mu}{\left(1 \mathord{-}\frac{\lambda_2}{\mu}\right)\left(1 \mathord{-}\frac{\lambda_1 \mathord{+}\lambda_2}{\mu}\right)}\mathord{+} 0.14\mathord{=}$
$0.33 \mathord{+} 0.625 \mathord{+} 0.14 \mathord{=}\SI{1.095}{\milli\second}<D^{\max}_1$, and $D_2 \mathord{=}$ $\frac{1/\mu}{\left(1 \mathord{-}\frac{\lambda_1}{\mu}\right)\left(1 \mathord{-}\frac{\lambda_1 \mathord{+}\lambda_2}{\mu}\right)}\mathord{+}\frac{1}{\mu \mathord{-}\lambda_2}$ $\mathord{=}0.83 \mathord{+}0.25 =\SI{1.08}{\milli\second}
<D^{\max}_2$.

\end{example}
\noindent\rule{1\columnwidth}{0.5mm}
In conclusion, the above example shows that no combination of per-service priorities results in both~$s_1$ and~$s_2$ meeting their target delays.
Instead moving one step up the flexibility ladder, i.e., assigning different priorities to service traffic flows across VNFs, allows the MNO to meet the vertical's requirements while increasing efficiency in resource usage, hence lowering the costs. As we will see later,  even better performance can be obtained through per-request priorities.

\section{System model and problem formulation}
\label{sec:model}

As mentioned earlier, the SDM-X has to use the VMs under its control to provide the newly-requested services with the required KPIs and at the minimum cost. Specifically, it has to make decisions on (i) {\em whether} existing VNF instances should to be shared; (ii) if so, {\em how} to assign the priorities to different services sharing the same VNF instance; (iii) {\em scaling up} the computational capabilities of the VMs if needed and possible.

As these decisions are made with reference to a single PoP, processing times are the dominant contribution with respect to network latency, and we therefore neglect the latter. This assumption is also justified by the ongoing work in the datacenter networking community (see, e.g.,~\cite{datacenter-survey}), where (i) switching is highly optimized, hence network delays tend to be very small, and (ii) network topologies tend to be highly regular, hence network delays between any two VMs tend to be very similar across pairs of VMs. It follows that network delays are often, {\em within a single PoP}, often negligible with respect to processing times.

Next, \Sec{sub-model} describes how all the entities that are involved in the VNF-sharing problem can be described,  while \Sec{problem}  defines the problem objective and constraints.

\subsection{System model}
\label{sec:sub-model}

The system model includes VMs~$m\in\Mc$, and VNFs~$v\in\Vc$~\footnote{VNFs can, in general, include multiple virtual deployment units (VDUs); without loss of generality,  we consider that each VNF includes only one VDU. Also, we assume that no VNF  requires isolation.}. Each VM runs (at most) one VNF, modeled as an M/M/1 queue with FIFO queuing and preemption, as widely assumed in recent works~\cite{placement-infocom,noi-infocom18,DBhamare17}. Also, let $C(m)$ be the maximum computation capability to which VM $m$ can be scaled up. We underline that, although in this work we focus on computational capability, memory and storage could be easily accommodated as well.  We refer to a VM as active if it hosts a VNF, and we express through binary variables~$y(v,m)$ whether VM~$m$ runs VNF~$v$. 

Different VNFs have different computational requirements, which are modeled through parameter~$l(v)$, expressing how many units of computational capability are needed to process one request in one second. A VNF with requirement~$l(v)=1$ running on a VM with capability~$\mu=1$ takes $l(v)/\mu=1$~time unit to process a service request. Using the same VM for a VNF with requirement~$l(v)=2$ yields a processing time of $l(v)/\mu=2$~time units. Notice how $l(v)$ values do not depend on the service using VNF~$v$.

Services~$s\in\Sc$ include one or more VNFs, and service~$s$ requests arrive at VNF~$v$ with rate~$\lambda(s,v)$; VNFs that are not used by a certain service have $\lambda$-values equal to~$0$. 
Through the~$\lambda(s,v)$ parameters, we can account for arbitrarily complex service (VNF) graphs where the number of requests can change between VNFs, and some requests may visit the same VNF more than once.
We focus on the maximum average delay~$D^{\max}(s)$ of  service~$s$ as target KPI, although our model can be seamlessly extended to account for different (or additional) KPIs, e.g., service request drop probability.

Each VM uses a quantity~$\mu(m)$ of computational capability, which can be scaled up till~$C(m)$. 
Then the rate at which a VNF~$v$ deployed at VM~$m$ processes service requests results to be~$\frac{\mu(m)}{l(v)}$. Finally, binary variables~$x(s,v,m)$ express whether service~$s$ uses the instance of VNF~$v$  at VM~$m$; this allows us to account for the fact that multiple instances of the same VNF can be deployed at different VMs. For clarity, the above parameters and variables are summarized in \Tab{notation}.

\begin{table}
\caption{Notation ($\dagger$~denotes variables of the modified problem described in \Sec{sub-assignment})
    \label{tab:notation}
} 
\scriptsize
\begin{tabularx}{\columnwidth}{|l|l|X|}
\hline
{\bf Symbol} & {\bf Type} & {\bf Meaning} \\
\hline
$C(m)$ & Parameter & Maximum capability to which VM~$m$ can be scaled up \\
\hline
$D^{\max}(s)$ & Parameter & Target delay for service~$s$ \\
\hline
$j$ & Parameter & Jitter applied when assigning per-request priorities \\
\hline
$l(v)$ & Parameter & Computational capability needed to process one request at VNF~$v$ \\
\hline
$\Mc=\{m\}$ & Set & Set of VMs \\
\hline
$\pi(s,v)$ & Random var. &
Describes the priority assigned to requests of service~$s$ upon entering VNF~$v$\\
\hline
$p(s,v)$ & Parameter & Per-VNF priority of service~$s$ at VNF~$v$ \\
\hline
$S(s,v)$ & Aux. var. & Sojourn time of requests of service~$s$ at VNF~$v$ \\
\hline
$\Sc=\{s\}$ & Set & Set of services \\
\hline
$\Vc=\{v\}$ & Set & Set of VNFs \\
\hline
$x(s,v,m)$ & Binary var. & Whether requests of service~$s$ use the instance of VNF~$v$ running at VM~$m$ \\
\hline
$y(v,m)$ & Binary var. & Whether VM~$m$ runs VNF~$v$ \\
\hline
$\kappa_f(m)$ & Parameter & Fixed cost incurred when activating VM~$m$ \\
\hline
$\kappa_p(m)$ & Parameter & Proportional cost incurred when using one unit of computational capability for VM~$m$ \\
\hline
$\Lambda(s,v)$ & Aux. var. & Arrival rate at VNF~$v$ of requests that are given priority over requests of service~$s$ \\
\hline
\rule{0pt}{7pt} \hspace{-1.5mm} $\tilde{\Lambda}(s,v)$ & \hspace{-1mm} Real var.$^\dagger$ & \hspace{-1mm}  Arrival rate at VNF~$v$ of requests that are given priority over requests of service~$s$\\
\hline
$\lambda(s,v)$ & Parameter & Rate at which requests of service~$s$ enter VNF~$v$ \\
\hline
$\mu(m)$ & Real var. & Computational capability to use for VM~$m$ \\
\hline
\end{tabularx}
\vspace*{-5mm}
\end{table}

\subsection{Problem formulation}
\label{sec:problem}

We now discuss the objective of the VNF-sharing problem and the constraints we need to honor.

{\bf Objective.}
The high-level goal of the MNO is to minimize the cost, which consists of two  components: a fixed cost~$\kappa_f(m)$ paid if  VM~$m$ is activated, and a proportional cost~$\kappa_p(m)$ paid for each unit of computational capability used therein. The objective is then given by:
\begin{equation}
\label{eq:obj}
\min_{y,\mu}\sum_{m\in\Mc}   \left(  \kappa_f(m)  \sum_{v\in\Vc} y(v,m) +\kappa_p(m)\mu(m)\right).
\end{equation}

{\bf Capability and instance limits.}
We must account for the 
maximum value~$C(m)$ to which the capability~$\mu(m)$ of each VM~$m$ can be scaled up: 
\begin{equation}
\label{eq:capacity}
\mu(m)\leq C(m)\quad \forall m\in\Mc.
\end{equation}
Also, we can have at most one VNF per VM:
\begin{equation}
\label{eq:y}
\sum_{v\in\Vc}y(v,m)\leq 1,\quad\forall m\in\Mc,
\end{equation}
and only active VMs can be used, i.e.,
\begin{equation}
\label{eq:only-existing}
y(v,m)\geq x(s,v,m),\quad\forall s\in\Sc,v\in\Vc, m\in\Mc.
\end{equation}

{\bf Service times.}
Each service~$s$ has a maximum average service time~$D^{\max}(s)$ that must be honored. Since, as discussed earlier, processing time is the dominant component of service time, this is equivalent to imposing:
\begin{equation}
\label{eq:max-service-time}
\sum_{v\in\Vc}S(s,v) \leq D^{\max}(s),\quad\forall s\in\Sc,
\end{equation}
where~$S(s,v)$ is the {\em sojourn time} (i.e., the time spent waiting or being served) experienced by requests of service~$s$ at VNF~$v$. By convention, we set~$S(s,v)=0$ if service~$s$ does not include VNF~$v$.

As detailed below, sojourn times, in turn, depend on:
\begin{itemize}
    \item the computational capability~$l(v)$ requested by the VNFs;
 \item the service request arrival rate at the VNFs~$\lambda(s,v)$;
    \item the priority of the service requests at the traversed VNFs;
    \item the computational capability~$\mu(m)$ assigned to the VM hosting the VNF instance processing the requests.
\end{itemize}
Using~\cite[Sec.~3.2]{kleinrock-vol2} and~\cite{dispensa}, we can generalize the expression used in \Ex{prios} and write the sojourn time of requests of service~$s$ at VNF~$v$, as:
\begin{equation}
\label{eq:S}
S(s,v){=}\frac{l(v)}{\mu(\bar{m})}\frac{1}{1-l(v)\frac{\Lambda(s,v)}{\mu(\bar{m})}}\frac{1}{1-l(v)\frac{\Lambda(s,v)+\lambda(s,v)}{\mu(\bar{m})}},
\end{equation}
where~$\bar{m}$ is the VM hosting the instance of VNF~$v$ used by service~$s$, i.e., $\bar{m}=m\in\Mc\colon x(s,v,m)=1$.

In \Eq{S}, $\Lambda(s,v)$ represents the arrival rate of requests (of any service) arriving at VNF~$v$ that are given a priority higher than a generic request of service~$s$.
Let~$\pi(s,v)$ be the random variable describing the priority assigned to requests of service~$s$ at VNF~$v$, then we have:
\begin{equation}
\label{eq:def-lambda}
\Lambda(s,v)=\sum_{t\in\Sc}\PP\left(\pi(t,v)>\pi(s,v)\right)\lambda(t,v).
\end{equation}
The intuitive meaning of \Eq{def-lambda} is that~$\Lambda(s,v)$ grows as it becomes more likely that requests of other services~$t\neq s$ are given higher priority over requests of service~$s$.

{\bf Problem complexity.} 
The actual expression of~$\Lambda(s,v)$ depends on the type of the $\pi(s,v)$~variables and is not guaranteed to be linear, convex, or even continuous.
It follows that, in the general case, the problem of setting the priorities in such a way to optimize \Eq{obj} has NP-hard complexity and is thus impractical for realistically-sized problem instances. Indeed, as will be more clear from \Sec{vlevel} and \Sec{rlevel}, finding the optimum in the per-request priority case, would require to search over all possible distributions of~$\pi(s,v)$.

Below, we compute~$\Lambda(s,v)$ for the relevant cases of per-VNF priorities and uniformly-distributed, per-request priorities.

\subsubsection{Per-VNF priorities}
\label{sec:vlevel}
We recall that, if per-VNF priorities are supported as in \Sec{example}, then all requests of each service~$s$ entering VNF~$v$ are given the same, deterministic priority. It follows that, denoted such a priority with~$p(s,v)$, in the per-VNF case  $\pi(s,v)$ is always distributed according to a Dirac delta function centered in~$p(s,v)$, i.e.,~$\delta\left(\pi(s,v)-p(s,v)\right)$. Hence, $\Lambda(s,v)$ is discontinuous and given by:
\begin{equation}
\label{eq:lambda-vlevel}
\Lambda(s,v)=\sum_{t\in\Sc}H\left (p(t,v)-p(s,v)\right )\lambda(t,v),
\end{equation}
where~$H(\cdot)$ is the Heaviside step function. Indeed, intuitively a request of service~$s$ will be queued after all requests of services~$t$ with higher priority than~$s$ (since~$H(p(t,v)-p(s,v))=1$ if~$p(t,v)>p(s,v)$), after half of the requests of services with the same priority as~$s$ (since~$H(p(t,v)-p(s,v))=0.5$ if~$p(t,v)=p(s,v)$), and before all other requests (since~$H(p(t,v)-p(s,v))=0$ if~$p(t,v)<p(s,v)$).

\subsubsection{Per-request priorities}
\label{sec:rlevel}
This case corresponds to higher flexibility and  implies that priorities could follow any distribution. Below, we focus on the simple, yet relevant, case where priorities are distributed uniformly between~$r(s,v)-j$ and~$r(s,v)+j$. In this case, let us define the quantity~$q(s,t,v)=\PP(\pi(t,v)>\pi(s,v))$, whose value can be computed   through the convolution of the pdfs of~$\pi(s,v)$ and~$\pi(t,v)$. Following the steps in~\cite{diffvar}, we get:
\begin{multline}
\label{eq:def-q}
\begin{aligned}
q(s,t,v) & = \PP(\pi(t,v)>\pi(s,v))\\
& = \PP\left(\pi(t,v)-\pi(s,v)>0\right)\\
\end{aligned}\\
=\begin{cases}
1 & \text{if } r(t,v){-}r(s,v)>2j\\
\frac{1}{2}{+} \frac{r(t,v)-r(s,v)}{4j} & \text{if } -2j {\leq} r(t,v){-}r(s,v){\leq} 2j\\
0 & \text{if } r(t,v){-}r(s,v)<-2j\,.
\end{cases} 
\end{multline}
Once the $q(s,t,v)$~are known, the $\Lambda(s,v)$ values can be computed by replacing \Eq{def-q} in \Eq{def-lambda}, obtaining:
\begin{equation}
\label{eq:def-lambda-rlevel}
\Lambda(s,v)=\sum_{t\in\Sc}q(s,t,v)\lambda(t,v).
\end{equation}
In this case, it is also possible to prove that the choice of parameter~$j$ has no impact on the solution space, hence, on the decisions that are made. 

\section{The FlexShare solution strategy}
\label{sec:algo}

\begin{figure}
\centering
\psfragfig[width=1\columnwidth]{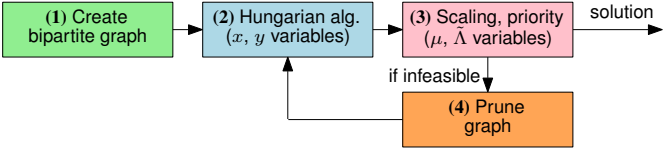}{
\psfrag{BG1}[c][m]{\scriptsize{{\sf {\bf (1)} Create}}}
\psfrag{BG2}[c][m]{\scriptsize{{\sf bipartite graph}}}
\psfrag{HU1}[c][m]{\scriptsize{{\sf {\bf (2)} Hungarian alg.}}}
\psfrag{HU2}[c][m]{\scriptsize{{\sf ($x$, $y$ variables)}}}
\psfrag{PR1}[c][m]{\scriptsize{{\sf {\bf (3)} Scaling, priority}}}
\psfrag{PR2}[c][m]{\scriptsize{{\sf ($\mu$, $\tilde{\Lambda}$ variables)}}}
\psfrag{PE1}[c][m]{\scriptsize{{\sf {\bf (4)} Prune}}}
\psfrag{PE2}[c][m]{\scriptsize{{\sf graph}}}
\psfrag{I1}[r][t]{\scriptsize{{\sf if infeasible}}}
\psfrag{I2}[r][m]{\scriptsize{{\sf if infeasible}}}
\psfrag{I3}[c][t]{\scriptsize{\hspace{5mm}{\sf solution}}}
} 
\caption{
    The FlexShare strategy. Step~1 builds a bipartite graph showing which VMs {\em could} run each VNF. Step~2 runs the Hungarian algorithm on such a graph, obtaining the optimal values for the~$x$- and~$y$-variables. Step~3 solves a convex variant of the original problem in \Sec{problem}. If feasible, its variables ($\mu$, $\tilde{\Lambda}$) are used to determine the scaling and the service priorities; otherwise, the bipartite graph is {\em pruned} (step~4) and the procedure restarts from step~2.
    \label{fig:flow}
} 
\vspace{-5mm}

\end{figure}

In light of the problem complexity, we propose  a fast, yet highly effective, solution strategy, named FlexShare, which runs iteratively and consists of four main steps, as outlined in \Fig{flow}. 
The first step, detailed in \Sec{sub-bipartite}, consists in building a bipartite graph including the VNFs to deploy, and the VMs that are active or can be activated. The edges of the graph express the possibility of using a VM to provide a VNF, either by sharing an existing instance or by deploying a new one. Edges are labeled with the cost associated with each decision, i.e., the~$\kappa_p$ and/or~$\kappa_f$ terms contributing to the objective \Eq{obj}. In step~2, the Hungarian algorithm~\cite{hungarian} is run on the generated bipartite graph, yielding the optimal, minimum-cost assignment of VNFs to VMs, i.e., the~$x$- and $y$-variables of the problem.

Given these decisions, step~3 aims at assigning the priorities and finding the amount of computational capability to use in every VM. To this end, a
simpler (namely, {\em convex}) variant of the problem defined in \Sec{problem} is formulated and solved, as detailed in \Sec{sub-assignment}.

If step~3 results in an infeasible problem, we {\em prune} the bipartite graph (step~4). The underlying intuition is that a cause for infeasibility is overly aggressive sharing of existing VNF instances. Therefore, as detailed in \Sec{sub-prune}, edges that result in an overload of VMs are removed from the bipartite graph. After pruning, the algorithm starts a new iteration with step~2. Moving from one iteration to the next means reducing the likelihood that VNF instances are shared between services, and thus increasing the cost incurred by the MNO, due to the~$\kappa_f$ fixed cost terms. The procedure stops as soon as one feasible solution is found.

\subsection{Steps 1--2: Bipartite graph and Hungarian algorithm}
\label{sec:sub-bipartite}

{\bf The bipartite graph.}
The purpose of the bipartite graph is to represent (i) the possible VNF assignment decisions, i.e., which VNFs can be provided at which VMs and which VNFs can be shared among services, and (ii) the associated cost incurred by the MNO.

More formally, the bipartite graph is created according to the following rules:
\begin{enumerate}
    \item a vertex is created for each VNF and for each VM;
    \item an edge is drawn from every VNF to every unused VM;
    \item an edge is drawn from every VNF to every VM currently running the same VNF, provided that the maximum computational capability of the VM is sufficient to guarantee stability.
\end{enumerate} 
Denoted by~$\bar{s}$ is the service now being deployed, VNF~$v$ can be provided at VM~$m$ while (still) ensuring stability if:
\begin{equation}
\label{eq:check-stability}
l(v)\sum_{s\in\Sc}\left [ \left(x(s,v,m)+\ind{s=\bar{s}}\right)\lambda(s,v)\right ]<C(m).
\end{equation}
The first member of \Eq{check-stability} is the total load imposed on VM~$m$ by services that are already being served therein, if any, {\em and} the new one~$\bar{s}$ (for which the indicator function is one). Through \Eq{check-stability}, we can then check that this quantity is lower than the maximum VM capability~$C(m)$.

Note that \Eq{check-stability} does {\em not} imply that~$\bar{s}$, or existing services, can be served {\em in time}, i.e., within their deadlines; indeed, this depends on the priority and computational capability assignment decisions, and cannot be checked at the graph generation time. The purpose of step~1 is just  to generate a graph accounting for all possible assignment options.

The cost of each edge connecting VM~$m$ with VNF~$v$ is given by the following expression:
\begin{equation}
\label{eq:edge-cost}
\left(1-\sum_{v}y(v,m)\right)\kappa_f(m)+\kappa_p(m)\left( l(v)\lambda(\bar{s},v)+\epsilon\right).
\end{equation}
In \Eq{edge-cost}, the first term is the fixed cost associated with activating VM~$m$, which is incurred only if $m$~is not already active (the summation can be at most~$1$, as per \Eq{y}). The second term is the proportional cost associated with the additional computation capability needed at VM~$m$ to guarantee stability, with~$\epsilon$ being a positive,  arbitrarily small value.

{\bf Hungarian algorithm and assignment decisions.}
The Hungarian algorithm~\cite{hungarian} is a combinatorial optimization algorithm with polynomial (cubic) time complexity in the number of edges in the graph. When applied to the bipartite graph we generate, it selects a subset of edges such that (i) each VNF is connected to exactly one VM, and (ii) the total cost of the selected edges is minimized.

Selected edges map to assignment decisions. Specifically, for each selected edge connecting VNF~$v$ and VM~$m$, we set $y(v,m)\gets 1$ and~$x(\bar{s},v,m)\gets 1$, i.e., we activate~$m$ (if not already active)  deploying therein an instance of~$v$,  and use it to serve service~$\bar{s}$. The obtained values for the~$x$ and $y$-variables are used in step~3 to decide priorities and computational capability assignment, as set out next.

\subsection{Step~3: Priority and scaling decisions}
\label{sec:sub-assignment}

The purpose of step~3 of the FlexShare procedure is to decide the priorities to assign to each VNF and service, as well as any needed scaling of VM computation capability. Since the complexity of the problem stated in \Sec{problem} depends on the
presence of the~$\pi(s,v)$ variables, we proceed as follows:
\begin{enumerate}
    \item we formulate a {\em simplified} problem, which contains no random variables and is guaranteed to be convex;
    \item we use the variables of the simplified problem to set the~$\mu(q)$ variables of the original problem, as well as the parameters of the distribution of the~$\pi(s,v)$ variables.
\end{enumerate}

{\bf Convex formulation.}
To avoid dealing with probability distributions, we replace the $\Lambda(s,v)$~auxiliary variables of the original problem with independent variables~$\tilde{\Lambda}(s,v)$, thus dispensing with \Eq{def-lambda}. Given $x$ and $y$, the decision variables of the modified problem are~$\tilde{\Lambda}(s,v)$ and~$\mu(m)$, while the objective is still given by \Eq{obj}. Having~$\tilde{\Lambda}(s,v)$ as a variable means deciding (intuitively) how many higher-priority service requests each incoming request will find. Such values are later mapped to the parameters of the distributions of~$\pi(s,v)$.

If we solve the modified problem with no further changes, the optimal solution would always yield~$\tilde{\Lambda}(s,v)=0,\forall s,v$, i.e., no request ever encounters higher-priority ones, which is clearly not realistic. To avoid that, we mandate that the average behavior, i.e., the average number of higher-priority requests met, is the same as in the original problem:
\begin{equation}
\label{eq:minlambda}
\sum_{s\in\Sc}\tilde{\Lambda}(s,v)=\frac{|\Sc|}{2}\sum_{s\in\Sc}\lambda(s,v),\quad\forall v\in\Vc.
\end{equation}
The intuition behind \Eq{minlambda} is that each $\Lambda(s,v)$-value (in the original problem) is the sum of several $\lambda$-values, i.e., the services arrival rates. The $\lambda$-value associated with the highest-priority service will contribute to~$|\Sc|-1$ $\Lambda(s,v)$-values, the one associated with the second-highest-priority service will contribute to $|\Sc|-2$ $\Lambda(s,v)$-values, and so on. On average, each $\lambda$-value contributes to~$\frac{|\Sc|}{2}$ $\Lambda(s,v)$-values.

It  can be proved that the modified problem is convex and, thus, solvable in polynomial time
(in the problem size, which depends on the number of VNFs and VMs)
through off-the-self, commercial solvers:
\begin{property}
\label{prop:convex}
The problem of minimizing \Eq{obj} subject to constraints \Eq{capacity}--\Eq{max-service-time} and \Eq{minlambda}, is convex.
\end{property}
See~\cite{proofs} for the proof.

{\bf Setting the variables of the original problem.}
After solving the convex problem described above, we can use the optimal solution thereof to make scaling decisions, i.e., to set the~$\mu(m)$ variables in the original problem, as well as  priority decisions, i.e., the parameters of the distribution of~$\pi(s,v)$. For $\mu(m)$, we can simply use the corresponding variables in the simplified problem, which have the same meaning and are subject to the same constraints. As for priorities, the procedure to follow depends on the type of priority adopted in the system at hand, hence, on the type of the variables~$\pi(s,v)$.

Specifically, if per-VNF priorities are supported (as in \Sec{vlevel}), we set the $p(s,v)$~values in such a way that services associated with a higher~$\Lambda(s,v)$ have lower priority, e.g., by imposing that $p(s,v)\gets-\tilde{\Lambda}(s,v)$.
If, on the other hand, we are in the case of \Sec{rlevel}, i.e., per-requests  priorities are uniformly distributed, then we can solve a system of linear equations where the~$\tilde{\Lambda}(s,v)$ from the solution of the simplified problem are known terms, the $q(s,t,v)$~quantities are the unknowns, and equations have the form \Eq{def-q} and \Eq{def-lambda-rlevel}.

Regardless the way priorities are assigned, it is important to stress that our approach has general validity and can be combined with {\em any type} of priority distribution.

\subsection{Step~4: graph pruning}
\label{sec:sub-prune}

If the problem we solve in step~3 (priority and scaling decisions) is infeasible, a possible cause lies in the decisions made in step~2  (the Hungarian algorithm), i.e., the~$x$ and $y$ variables. Therefore, we restart from step~2  considering a different bipartite graph, more likely to result in a feasible problem.

To this end, we consider the {\em irreducible infeasible set} (IIS) of the problem instance solved in step 3, i.e., the set of constraints therein that, if removed, would yield a feasible problem. Given the IIS, we proceed as follows:
\begin{enumerate}
    \item we identify constraints in the IIS of type \Eq{capacity}, thus, a set of VMs that would need more capability;
    \item among such VMs, we select those that are used by the newly-deployed service~$\bar{s}$;
    \item among them, we identify the one that is the closest to instability, i.e., the VM~$m^\star$ minimizing the quantity~$C(m)-\sum_{s\in\Sc} x(s,v^\star,m)l(v^\star)\lambda(s,v^\star)$, where $v^\star$~is the VNF deployed at~$m$;
    \item we prune from the bipartite graph the edge between~$v^\star$ and~$m^\star$.
\end{enumerate}
The intuitive reason for this procedure is that a cause for delay constraints violations is that the newly-deployed service~$\bar{s}$ is causing one of the VMs it uses to operate too close to instability, and thus with high delays. By removing the corresponding edge from the bipartite graph, we ensure that VM~$m^\star$ is not used by service~$\bar{s}$.

Note that we are guaranteed that the IIS contains at least one constraint of type \Eq{capacity} thanks to the following result, proved in~\cite{proofs}:
\begin{theorem}
\label{thm:iis}
Every infeasible instance of the modified problem presented in \Sec{sub-assignment} includes at least one constraint of type \Eq{capacity} in its IIS.
\end{theorem}
FlexShare then restarts with step~2, where the Hungarian algorithm takes as an input the pruned bipartite graph.

\subsection{Computational complexity}
\label{sec:polynomial}

The FlexShare strategy has polynomial worst-case computational complexity. Specifically:
\begin{itemize}
    \item step~1 involves a simple check over at most~$|\Vc||\Mc|$ VNF/VM pairs;
    \item step~2, the Hungarian algorithm, has cubic complexity in the number of nodes in the graph~\cite{hungarian};
    \item step~3 requires solving a convex problem, as proven in \Prop{convex}, and the resulting complexity is also cubic;
    \item step~4 iterates over at most~$|\Mc|$ constraints of type \Eq{capacity}, and thus it has linear complexity;
    \item the whole procedure is repeated for (at most) as many times as there are edges in the original bipartite graph.
\end{itemize}
\begin{figure}
\centering
\psfragfig[width=.6\columnwidth]{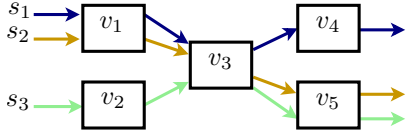}{
\psfrag{v1}[c][m]{$v_1$}
\psfrag{v2}[c][m]{$v_2$}
\psfrag{v3}[c][m]{$v_3$}
\psfrag{v4}[c][m]{$v_4$}
\psfrag{v5}[c][m]{$v_5$}
\psfrag{s1}[c][b]{$s_1$}
\psfrag{s2}[c][b]{$s_2$}
\psfrag{s3}[c][b]{$s_3$}
} 
\caption{
    VNF graphs in the synthetic scenario.
    \label{fig:synth}
} 
\vspace{-5mm}
\end{figure}

\begin{figure*}
\centering
\includegraphics[width=.3\textwidth]{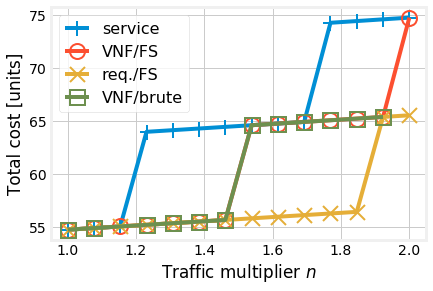}
\includegraphics[width=.3\textwidth]{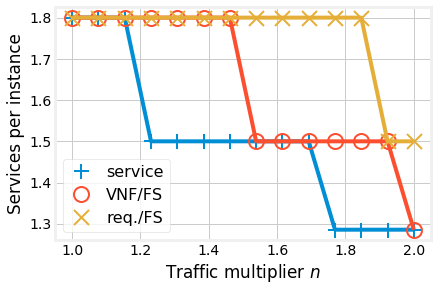}
\includegraphics[width=.3\textwidth]{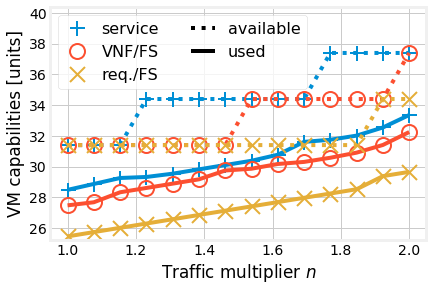}
\caption{
Synthetic scenario: total cost (left); average number of services sharing a VNF instance (center); used and maximum VM capability (right).
    \label{fig:synth-details}
} 
\vspace{-4mm}
\end{figure*}

\section{Numerical results}
\label{sec:results}

In this section, we describe the reference scenarios and benchmark solutions we consider, in \Sec{scenario}; then we present numerical results obtained under the synthetic and realistic scenarios in, respectively, \Sec{res-synth} and \Sec{res-real}.

\subsection{Reference scenarios and benchmarks}
\label{sec:scenario}

{\bf Synthetic scenario.}
It includes three services~$s_1\dots s_3$, sharing five VNFs~$v_1\dots v_5$, as depicted in \Fig{synth}. All VNFs have coefficient~$l(v)=10^{-3}$~units/request, while the arrival rates associated with each service vary between~1 and \SI{2}{requests/\milli\second}. Target delays range between \SI{20}{\milli\second} for~$s_1$ and~\SI{5}{\milli\second} for~$s_3$. The scenario includes 10~VMs, whose fixed and proportional costs are~$\kappa_f=8$ and~$\kappa_p=0.5$ units, respectively, and whose capability is randomly distributed between~5 and 10~units. Such a scenario is small enough to allow a comparison against optimal priority assignments found by brute-force. At the same time, it contains many interesting features, including different combinations of services sharing different VNFs and different cost/capability trade-offs at VMs.

\begin{figure*}
\centering

\includegraphics[width=.3\textwidth]{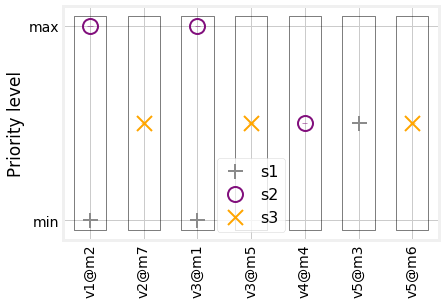}
\includegraphics[width=.3\textwidth]{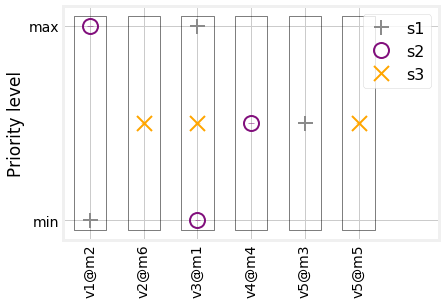}
\includegraphics[width=.3\textwidth]{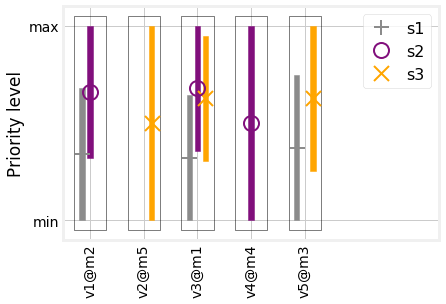}
\caption{
\vspace*{-2mm}
Synthetic scenario, $n=1.8$: priorities assigned to each service with per-service (left), per-VNF (center), and per-request priorities (right).
    \label{fig:synth-prios}
} 
\vspace{-5mm}
\end{figure*}

{\bf Realistic scenario.}
We consider three services, all connected to the smart-city domain:
\begin{itemize}
    \item intersection collision avoidance (ICA): vehicles periodically broadcast a message (e.g., CAM) including their position and speed; a collision detector checks if any pair of them are on a collision course and, if so, it issues an alert;
    \item vehicular see-through (CT): cars display on their on-board screen the video captured by the preceding vehicle, e.g., a large truck obstructing the view;
    \item urban sensing based on the Internet-of-Things (IoT).
\end{itemize}
\Tab{lambdas}, based on~\cite{pimrc-wp3,mtc}, reports the VNFs used by each service and the associated arrival rates. All services share the EPC child service, which is itself composed of five VNFs. Furthermore, the car information management (CIM) database can be shared between the ICA and the CT services.

We leverage the real-world mobility trace~\cite{lust},  to  assign user density and request rates. Focusing on a downtown intersection, we consider that (i) all vehicles within~50~m from the intersection are users of the ICA service, and send a CAM  every~0.1~s; (ii) all vehicles within~100~m from the intersection are users of the CT service, and send a request (i.e., refresh their video) every~200~ms; (iii) a total of 200~sensors are deployed in the area, each generating, according to the traffic model described in the 3GPP standard~\cite{3gppmtc}, one request every~0.1~s.
\Tab{lambdas} reports the resulting request rates and the requirement~$l(v)$ associated with each VNF. As discussed in \Sec{model}, the $\lambda(s,v)$ values also incorporate the fact that not all requests visit all VNFs of a service, e.g., all ICA requests visit the collision detector but only one in ten needs the alarm generator. 

Finally, we assume that the PoP contains 10~VMs, each of which can be scaled up to at most~$C(m)=\SI{1000}{units}$, and each associated with fixed and proportional costs of~$\kappa_f=\SI{1000}{units}$ and $\kappa_p=\SI{1}{unit}$, respectively.

\begin{table}
\caption{
Realistic scenario: request rate and computational load associated with every VNF
    \label{tab:lambdas}
} 
\scriptsize{
\begin{tabularx}{\columnwidth}{|X|r|r|}
\hline
{\bf VNF} & {\bf Rate~$\lambda(s)$} & {\bf Requirement~$l(v)$}\\
\hline\hline
\multicolumn{3}{|>{\hsize=\columnwidth}c|}{{\em Intersection collision avoidance (ICA)}}\\
\hline
eNB & 117.69 & $10^{-4}$\\
\hline
EPC PGW & 117.69 & $10^{-4}$\\
\hline
EPC SGW & 117.69 & $10^{-4}$\\
\hline
EPC HSS & 11.77 & $10^{-4}$\\
\hline
EPC MME & 11.77 & $10^{-3}$\\
\hline
Car information management (CIM) & 117.69 & $10^{-3}$\\
\hline
Collision detector & 117.69 & $10^{-3}$\\
\hline
Car manufacturer database & 117.69 & $10^{-4}$\\
\hline
Alarm generator & 11.77 & $10^{-4}$\\
\hline
\multicolumn{3}{|>{\hsize=\columnwidth}c|}{{\em See through (CT)}}\\
\hline
eNB & 179.82 & $10^{-4}$\\
\hline
EPC PGW & 179.82 & $10^{-4}$\\
\hline
EPC SGW & 179.82 & $10^{-4}$\\
\hline
EPC HSS & 17.98 & $10^{-4}$\\
\hline
EPC MME & 17.98 & $10^{-3}$\\
\hline
Car information management (CIM) & 179.82 & $10^{-3}$\\
\hline
CT server & 179.82 & 5 $10^{-3}$\\
\hline
CT database & 17.98 & $10^{-4}$\\
\hline
\multicolumn{3}{|>{\hsize=\columnwidth}c|}{{\em Sensing (IoT)}}\\
\hline
eNB & 50 & $10^{-4}$\\
\hline
EPC PGW & 50 &$10^{-4}$\\
\hline
EPC SGW & 50 & $10^{-4}$\\
\hline
EPC HSS & 5 & $10^{-4}$\\
\hline
EPC MME & 5 & $10^{-3}$\\
\hline
IoT authentication & 20 & $10^{-4}$\\
\hline
IoT application server & 20 & $10^{-3}$\\
\hline
\end{tabularx}
} 
\vspace{-5mm}
\end{table}

{\bf Benchmark solutions.}
We study the performance of the following strategies, in increasing order of flexibility:
\begin{itemize}
    \item {\bf service}: per-service priorities, with lowest-delay services having the highest priority;
    \item {\bf VNF/FS}: per-VNF priorities, assigned with FlexShare;
    \item {\bf VNF/brute}: optimal per-VNF priorities (with brute-force);
    \item {\bf req/FS}: per-request, uniformly distributed priorities, assigned with FlexShare.
\end{itemize}

\subsection{Results: synthetic scenario}
\label{sec:res-synth}

We start by considering the synthetic scenario and, in order to study different traffic conditions, multiply the arrival rates by a factor~$n$ ranging between~1 and~2.

\Fig{synth-details}(left) focuses on the main metric we consider, namely, the total cost incurred by the MNO. We can observe that, as one might expect,  higher traffic translates into higher cost. More importantly, more flexibility in priority assignment results in substantial cost savings. As for per-VNF priorities, they exhibit an intermediate behavior between per-service and per-request ones, with virtually no difference between the case where FlexShare is used to determine the priorities (``VNF/FS'') and that where all possible options are tried out in a brute-force fashion (``VNF/brute''). This highlights the effectiveness of the FlexShare strategy, which can make optimal decisions in almost all cases with low complexity.

\Fig{synth-details}(center) shows the average number of services sharing a VNF instance. It is clear that a higher flexibility in priority assignment results in more sharing, hence fewer VNF instances deployed. As $n$~increases, the number of services per instance decreases: scaling up (i.e., increasing the capability of VMs) is insufficient, and scaling out (i.e., increasing the number of instances) becomes necessary.

This is confirmed by \Fig{synth-details}(right), depicting the total used VM capability (i.e.,~$\sum_{m\in\Mc}\mu(m)$) as well as the sum of the maximum values to which the capability of active VMs can be scaled up (i.e.,~$\sum_{m\in\Mc}C(m)y(m)$), denoted by solid and dotted lines, respectively. Both quantities grow with~$n$ and decrease as flexibility becomes higher. This makes intuitive sense for the maximum capability: \Fig{synth-details}(left) shows that under higher-flexibility strategies fewer VNF instances, hence fewer active VMs, are needed. Importantly, {\em used} capability values, i.e.,~$\mu(m)$, also decrease with flexibility. Indeed, higher flexibility makes it easier to match the computational capability obtained by each service within each VNF, with its needs.

\Fig{synth-prios} provides a qualitative view of how priorities are assigned to different services across different VNF instances. When priorities are assigned on a per-service basis (\Fig{synth-prios}(left)), services with lower target delay invariably have higher priority. If priorities are assigned on a per-VNF basis, as in \Fig{synth-prios}(center), the priorities of different services can change across VNF instances, e.g., $s_2$~has priority over~$s_1$ in the $v_1$ instance deployed at VM~$m_2$, but the opposite happens in the $v_3$ instance deployed at VM~$m_1$. \Fig{synth-prios}(right) shows that if per-request priorities are possible, services can be combined in any way at each VNF instance.

\subsection{Results: realistic scenario}
\label{sec:res-real}

\begin{figure*}
\centering
\includegraphics[width=.3\textwidth]{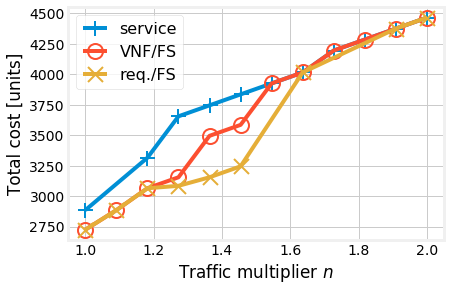}
\includegraphics[width=.3\textwidth]{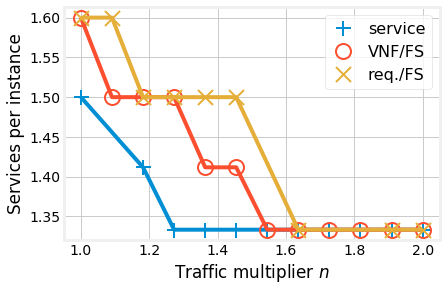}
\includegraphics[width=.3\textwidth]{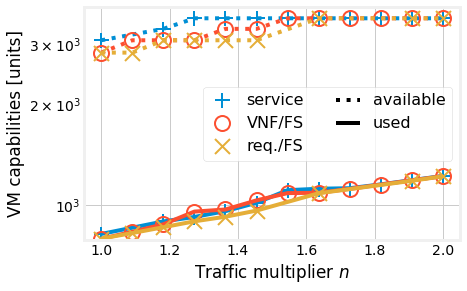}
\caption{
Realistic scenario: total cost (left); average number of services sharing a VNF instance (center); used and maximum VM capability (right).
    \label{fig:real-details}
} 
\vspace{-4mm}
\end{figure*}

We now move to the realistic scenario, again multiplying the arrival rates reported in \Tab{lambdas} by a factor~$n$ varying between~1 and~2. Recall that, owing to the larger scenario size, no comparison with the brute-force strategy is possible.

\Fig{real-details}(left) shows how the total cost yielded by the different strategies has the same behavior as in the synthetic scenario (\Fig{synth-details}(left)): the higher the flexibility, the lower the cost. Furthermore, for very high values of~$n$, all strategies yield the same cost; in those cases,
few or no VNF instances can be shared, regardless of how priorities are assigned.

\Fig{real-details}(center) shows that VNF instances are shared among services; by comparing it to \Fig{synth-details}(center) we can observe how the behavior of per-VNF priorities tends to be closer to per-request priorities than to per-service ones. This suggests that, even in large and/or complex scenarios, per-VNF priorities can be a good compromise between performance and implementation complexity.

\Fig{real-details}(right) shows a much larger difference between used and maximum capabilities compared to \Fig{synth-details}(right). This is due to the fact that, as can be seen from \Tab{lambdas}, there are fewer VNFs that are common among different services, and thus fewer opportunities for sharing.

\section{Related work}
\label{sec:relwork}

5G networks based on network slicing have attracted substantial attention, with several works focusing on 5G architecture~\cite{slicing1,slicing2}, associated decision-making issues~\cite{slicingp2}, and security~\cite{orch-sec}.

As one of the most important decisions to make in 5G environments, VNF placement has been the focus of several studies. One popular approach is optimizing a network-centric metric, e.g., load balancing~\cite{AHirwe16} or network utilization~\cite{TKuo16}. Other papers use cost functions, e.g.,~\cite{MMechtri16}, possibly including energy-efficiency considerations~\cite{AMarotta16}.

The aforementioned works typically result in mixed-integer linear programming (MILP) models. Other works cast VNF placement into a generalized assignment~\cite{infocom15_optimal}
or a set cover problem~\cite{eff-algos}.

{\bf Novelty.}
A first novel aspect of our work is the problem we consider, i.e., VNF-sharing within one PoP as opposed to traditional VNF placement.
From the modeling viewpoint, we depart from existing works in three main ways: (i) priorities are used as a decision variable rather than as an input; (ii) different priority-assignment schemes with different flexibility are accounted for and compared; (iii) the relationship between the amount of computational resources assigned to VNFs and their performance is modeled and studied;
(iv) VM capacity scaling is properly accounted for as a necessary, complementary aspect of VNF sharing.

\section{Conclusion}
\label{sec:conclusion}

We considered the problem of sharing VNFs among 5G services using the same PoP, and identified in service priority management one of its key aspects. We then proposed a solution strategy called FlexShare, able to efficiently make high-quality priority and VM scaling decisions. Our performance evaluation has shown that higher flexibility in priority assignment yields lower costs, and that FlexShare is able to provide near-optimal performance in all scenarios.

Future work will focus on
formally characterizing the performance gap between FlexShare and the optimal solution. Furthermore, with reference to the per-request priority case,
we will study additional distributions and assess their impact  on the resulting performance.

\bibliographystyle{IEEEtran}
\bibliography{refs}

\begin{thebibliography}{10}
\providecommand{\url}[1]{#1}
\csname url@samestyle\endcsname
\providecommand{\newblock}{\relax}
\providecommand{\bibinfo}[2]{#2}
\providecommand{\BIBentrySTDinterwordspacing}{\spaceskip=0pt\relax}
\providecommand{\BIBentryALTinterwordstretchfactor}{4}
\providecommand{\BIBentryALTinterwordspacing}{\spaceskip=\fontdimen2\font plus
\BIBentryALTinterwordstretchfactor\fontdimen3\font minus
  \fontdimen4\font\relax}
\providecommand{\BIBforeignlanguage}[2]{{%
\expandafter\ifx\csname l@#1\endcsname\relax
\typeout{** WARNING: IEEEtran.bst: No hyphenation pattern has been}%
\typeout{** loaded for the language `#1'. Using the pattern for}%
\typeout{** the default language instead.}%
\else
\language=\csname l@#1\endcsname
\fi
#2}}
\providecommand{\BIBdecl}{\relax}
\BIBdecl

\bibitem{slicing}
{NGMN Alliance}, ``Description of network slicing concept,'' 2016.

\bibitem{slicing2}
P.~Rost \emph{et~al.}, ``{Network slicing to enable scalability and flexibility
  in 5G mobile networks},'' \emph{IEEE Comm. Mag.}, 2017.

\bibitem{5gppp-architecture}
{5G PPP Architecture Working Group}. (2017) {View on 5G Architecture}.

\bibitem{ietf-mano}
{IETF}. (2017) {Network Slicing Management and Orchestration}.

\bibitem{placement-cinesi}
J.~Cao \emph{et~al.}, ``{VNF placement in hybrid NFV environment: Modeling and
  genetic algorithms},'' in \emph{{IEEE ICPADS}}.

\bibitem{placement-infocom}
R.~Cohen \emph{et~al.}, ``Near optimal placement of virtual network
  functions,'' in \emph{{IEEE INFOCOM}}, 2015.

\bibitem{noi-infocom18}
S.~Agarwal \emph{et~al.}, ``{Joint VNF Placement and CPU Allocation in 5G},''
  in \emph{IEEE INFOCOM}, 2018.

\bibitem{etsimano}
{ETSI}. (2017) {Network Functions Virtualisation (NFV); Management and
  Orchestration}.

\bibitem{pimrc-wp4}
K.~Antevski \emph{et~al.}, ``Resource orchestration of {5G} transport networks
  for vertical industries,'' in \emph{IEEE PIMRC}, 2018.

\bibitem{norma}
B.~Sayadi \emph{et~al.}, ``{SDN for 5G Mobile Networks: NORMA perspective},''
  in \emph{Springer CROWNCOM}, 2016.

\bibitem{ngmn-mano}
{NGMN Alliance}. (2017) {5G Network and Service Management including
  Orchestration}.

\bibitem{kleinrock-vol2}
L.~Kleinrock, \emph{Queueing systems: Computer applications}.\hskip 1em plus
  0.5em minus 0.4em\relax John Wiley \& Sons, 1976.

\bibitem{datacenter-survey}
W.~Xia \emph{et~al.}, ``{A survey on data center networking (DCN):
  Infrastructure and operations},'' \emph{IEEE Comm. surveys \& tutorials},
  2017.

\bibitem{DBhamare17}
D.~Bhamare \emph{et~al.}, ``Optimal virtual network function placement in
  multi-cloud service function chaining architecture,'' \emph{Computer
  Communications}, 2017.

\bibitem{dispensa}
{Malathi Veeraraghavan}. (2014) {Priority queueing}.
  \url{http://www.ece.virginia.edu/mv/edu/715/lectures/PQ.pdf}.

\bibitem{diffvar}
{Dimitrios Milios}. (2009) {Probability Distributions as Program Variables}.
  \url{http://www.inf.ed.ac.uk/publications/thesis/online/IM090722.pdf}.

\bibitem{hungarian}
H.~W. Kuhn, ``The hungarian method for the assignment problem,'' \emph{Wiley
  Naval Research Logistics}, 1955.

\bibitem{proofs}
{Proofs}. \url{https://dl.dropbox.com/s/9ylmq71iyyuqjrk/proofs.pdf}.

\bibitem{pimrc-wp3}
C.~Casetti \emph{et~al.}, ``Arbitration among vertical services,'' in
  \emph{IEEE PIMRC}, 2018, \url{http://arxiv.org/abs/1807.11196 }.

\bibitem{mtc}
T.~Taleb, A.~Ksentini, and A.~Kobbane, ``Lightweight mobile core networks for
  machine type communications,'' \emph{IEEE Access}, 2014.

\bibitem{lust}
L.~Codeca, R.~Frank, and T.~Engel, ``{Luxembourg SUMO Traffic (LuST) Scenario:
  24 hours of mobility for vehicular networking research},'' in \emph{IEEE
  VNC}, 2015.

\bibitem{3gppmtc}
{3rd Generation Partnership Project}, ``{3GPP} specification: 37.868; {RAN}
  improvements for machine-type communications,'' Tech. Rep.

\bibitem{slicing1}
H.~Zhang \emph{et~al.}, ``{Network slicing based 5G and future mobile networks:
  mobility, resource management, and challenges},'' \emph{IEEE Comm. Mag.},
  2017.

\bibitem{slicingp2}
K.~Samdanis \emph{et~al.}, ``{5G Network Slicing -- Part 2: Algorithms and
  practice},'' \emph{IEEE Comm. Mag.}, 2017.

\bibitem{orch-sec}
M.~A.~S. Santos \emph{et~al.}, \emph{Security Requirements for Multi-operator
  Virtualized Network and Service Orchestration for 5G}.\hskip 1em plus 0.5em
  minus 0.4em\relax Springer, 2017.

\bibitem{AHirwe16}
A.~Hirwe and K.~Kataoka, ``{LightChain: A lightweight optimization of VNF
  placement for service chaining in NFV},'' in \emph{IEEE NetSoft}, 2016.

\bibitem{TKuo16}
T.~W. Kuo \emph{et~al.}, ``Deploying chains of virtual network functions: On
  the relation between link and server usage,'' in \emph{IEEE INFOCOM}, 2016.

\bibitem{MMechtri16}
M.~Mechtri, C.~Ghribi, and D.~Zeghlache, ``A scalable algorithm for the
  placement of service function chains,'' \emph{IEEE Trans. on Network and
  Service Management}, 2016.

\bibitem{AMarotta16}
A.~Marotta and A.~Kassler, ``A power efficient and robust virtual network
  functions placement problem,'' in \emph{IEEE ITC}, 2016.

\bibitem{infocom15_optimal}
R.~Cohen \emph{et~al.}, ``{Near optimal placement of virtual network
  functions},'' in \emph{{IEEE INFOCOM}}, 2015.

\bibitem{eff-algos}
A.~Tomassilli \emph{et~al.}, ``Provably efficient algorithms for placement of
  service function chains with ordering constraints,'' in \emph{IEEE INFOCOM},
  2018.

\end{thebibliography}

\section*{Acknowledgment}

This work has been performed in the framework of the European Union's Horizon 2020 projects 5G-EVE and 5G-CARMEN, co-funded by the EU under grant agreements (respectively) No. 815074 and No. 825012. The views expressed are those of the authors and do not necessarily represent the project. The Commission is not liable for any use that may be made of any of the information contained therein.

\setcounter{property}{0}
\setcounter{theorem}{0}

\clearpage
\pagenumbering{gobble}

\appendix
\section*{Proofs}

\begin{property}
The problem of minimizing \Eq{obj} subject to constraints \Eq{capacity}--\Eq{max-service-time} and \Eq{minlambda}, is convex.
\end{property}
\begin{IEEEproof}
For the problem to be convex, the objective and all constraints must be so. Our expressions are linear, and thus convex. However, \Eq{max-service-time} contains $S(s,v)$-terms, which have to be proven to be convex. We do so by computing the second derivative of the expression~$S(s,v)$ in the~$\mu(m)$ and~$\Lambda(s,v)$ variables. It is easy to verify that, since the quantities $\tilde{\Lambda}(s,v)$, $\mu(m)$, $\lambda(s,v)$ and  $l(v)$ are all positive and  the system is stable (i.e.,  $l(v)\Lambda(s,v)<\mu(m)$ and~$l(v)(\Lambda(s,v)+\lambda(s,v))<\mu(m)$),  both derivatives are positive, which proves the thesis.
\end{IEEEproof}

\begin{theorem}
Every infeasible instance of the modified problem presented in \Sec{sub-assignment} includes at least one constraint of type \Eq{capacity} in its IIS.
\end{theorem}
\begin{IEEEproof}
The constraints of the modified problem are of type \Eq{capacity}--\Eq{max-service-time} and \Eq{minlambda}. Proving that there is a constraint of type \Eq{capacity} in the IIS is equivalent to proving that we can solve a violation of the other types of constraint by violating one or more constraints of type \Eq{capacity}.
Indeed, if a max-delay constraint of type \Eq{max-service-time} is violated, we can make the capacity of the VNF used by that service arbitrarily high; so doing, we can solve the violation of \Eq{max-service-time} at the cost of  violating \Eq{capacity}. 
Similarly, solving a violation of \Eq{minlambda} requires increasing the $\tilde{\Lambda}$-values, which in turn increases the sojourn times and results in a violation of \Eq{max-service-time}-type constraints, thus reducing to the previous case.
\end{IEEEproof}

\end{document}